# Control of hidden ground state order in NdNiO$_3$ superlattices

Ankit S. Disa[1,2], Alexandru B. Georgescu[1,2], James L. Hart[3], Divine P. Kumah[1,2], Padraic Shafer[4], Elke Arenholz[4], Dario A. Arena[4], Sohrab Ismail-Beigi[1,2], Mitra L. Taheri[3], Frederick J. Walker[1,2], Charles H. Ahn[1,2*]

[1]Center for Research on Interface Structures and Phenomena, Yale University, New Haven, Connecticut 06511, USA
[2]Department of Applied Physics, Yale University, New Haven, Connecticut 06511, USA
[3]Department of Materials Science & Engineering, Drexel University, Philadelphia, Pennsylvania 19104, USA
[4]Advanced Light Source, Lawrence Berkeley National Laboratory, Berkeley, California 94720, USA
[4]Department of Physics, University of South Florida, Tampa, FL 33620, USA

**Abstract:**

The combination of charge and spin degrees of freedom with electronic correlations in condensed matter systems leads to a rich array of phenomena, such as magnetism, superconductivity, and novel conduction mechanisms. While such phenomena are observed in bulk materials, a richer array of behaviors becomes possible when these degrees of freedom are controlled in atomically layered heterostructures, where one can constrain dimensionality and impose interfacial boundary conditions. Here, we unlock a host of unique, hidden electronic and magnetic phase transitions in NdNiO$_3$ while approaching the two-dimensional (2D) limit, resulting from the differing influences of dimensional confinement and interfacial coupling. Most notably, we discover a new phase in fully 2D, single layer NdNiO$_3$, in which all signatures of the bulk magnetic and charge ordering are found to vanish. In addition, for quasi two-dimensional layers down to a thickness of two unit cells, bulk-type ordering persists but separates from the onset of insulating behavior in a manner distinct from that found in the bulk or thin film nickelates. Using resonant x-ray spectroscopies, first-principles theory, and model calculations, we propose that the single layer phase suppression results from a new mechanism of interfacial electronic reconstruction based on ionicity differences across the interface, while the phase separation in multi-layer NdNiO$_3$ emerges due to enhanced 2D fluctuations. These findings provide insights into the intertwined mechanisms of charge and spin ordering in strongly correlated systems in reduced dimensions and illustrate the ability to use atomic layering to access hidden phases.

## I. Introduction

Strong electronic correlations result in a diverse array of practical and novel solid state phenomena. Predictively engineering electronically correlated materials has not yet reached the level realized for semiconductors, where almost all the electronic behavior is described accurately by single particle band theory. Over the past several years, the ability to control the synthesis of atomically precise, digital, complex oxide heterostructures has facilitated the study of strongly correlated electron systems in confined geometries [1]. Such research has led to the discovery of emergent electronic, magnetic, and orbital phases in artificially layered materials that are not accessible in their bulk counterparts [2]. Well-known examples of the unique properties arising in multi-component oxide heterostructures include interfacial ferromagnetism [3], improper ferroelectricity [4], polar metals [5], multiferroics [6], and enhanced superconductivity [7]. Generally, these effects occur from a combination of the dimensional confinement and the coupling of structure, charge, and spin imposed by the interface. As we show in this work, extensive phase behavior concealed in regions of the bulk phase diagram of correlated oxides can be unlocked by atomically tuning the relative influence of these two mechanisms.

The rare-earth nickelates ($R$NiO$_3$) have provided an exemplary model platform for investigating correlated phenomena because of the tunability of the phase transitions present in the system [8,9]. The metal-insulator transition temperature ($T_{MI}$) can be changed by choice of rare-earth ion ($R$) from ~600K (Lu) to ~0K (La) and can be distinct from or coincident with the Néel temperature ($T_N$) of the paramagnetic to antiferromagnetic transition (for Pr and Nd) [10]. Pressure, chemical doping, electric field, and strain have also been used to modulate the transitions [11-13]. Disentangling the driving force of the transitions is an ongoing discussion, and several potential models have been put forth, including those based on Mott physics [14], bandwidth control [10,15], Fermi surface instabilities [16,17], charge/bond ordering [18,19], and a combination of mechanisms [20].

Recent work on nickelate superlattices, spurred by predictions of potential superconductivity [21,22], showed that phase transitions [23-25] and electrical properties [26-28] can be controlled by dimensionality and interfacial constraints. It was also found that the orbital states can be strongly modified [29-32] and that magnetism can be induced [33-35] at nickelate interfaces. These results have shed light on methods to understand and control collective ordering in correlated systems using nickelate superlattices as an ideal testing ground. However, many questions remain regarding the relative sensitivity of such systems to various imposed boundary conditions. The study presented here addresses these questions by considering superlattices of NdNiO$_3$, a material whose properties lie close to the boundary between coincident and separated transitions, allowing one to separate the influence of dimensional confinement and electronic interfacial coupling.

In particular, we examine the evolution of electronic and magnetic states in atomically layered NdNiO$_3$/NdAlO$_3$ superlattices as a function of the NdNiO$_3$ thickness by means of electrical transport, x-ray absorption, and resonant soft x-ray scattering. The key results demonstrate new, hidden phases that arise as the thickness is reduced to the 2D limit: for layer thicknesses of 2 and 4 unit cells (uc), $T_{MI}$ separates substantially from $T_N$ and $T_{CO}$ (charge ordering temperature) whereas single, 2D NdNiO$_3$ layers show insulating behavior with no observable indications of electronic or spin ordering. Through first-principles and model theoretical calculations, it is ascertained that the origin of the suppression of ground state ordering is related to a distinctive electronic and orbital reconstruction at the interface, while the separation of $T_{MI}$ and $T_N$/$T_{CO}$ can be traced to reduced dimensionality. The observed phases have no bulk counterparts, and the fully 2D heterostructure represents an important, previously unexplored system for understanding correlated electron behavior in reduced dimensions. In addition, the mechanism driving the interfacial reconstruction in the single layer NdNiO$_3$ superlattice is novel and general – based on the relative ionicity of the cations – and can be potentially transferred to engineer orbital states in other correlated oxides.

The paper is organized as follows. First, we detail the sample preparation (Sec. II) and experimental measurements (Sec. III). In Sec. IV, we present the theoretical calculations and consider the nature of the interfacial reconstruction. We further contrast our observations with previous reports on nickelate superlattices, and discuss the implications of our findings in the context of correlated models of the phase transitions. We also remark on alternative possibilities for the observed phase behavior.

## II. Sample Details and Theoretical Methods

We grow superlattices of (NdNiO$_3$)$_m$/(NdAlO$_3$)$_4$ on LaAlO$_3$ (001) substrates using oxygen plasma assisted molecular beam epitaxy (MBE), following the procedure detailed in Ref. [9]. The NdNiO$_3$ layers in the superlattices are separated by insulating NdAlO$_3$ spacer layers four unit cells thick to inhibit interlayer coupling. During growth, the substrate temperature is maintained at $T$ = 590°C with an oxygen partial pressure of $P_{O_2}$ = 5-6 x 10$^{-6}$ Torr. A radio-frequency oxygen plasma source provides activated oxygen to the sample. Each layer is grown by co-deposition with a 1:1 flux ratio of cations as determined by quartz crystal microbalance measurements prior to growth. The thickness of the NdNiO$_3$ layer is varied from $m$ = 4, 2, 1 unit cells (with 11, 16 and 20 repeats of the superlattice, respectively) to explore the effects of dimensionality and confinement on the properties of the NdNiO$_3$ layers. For comparison, we grow a 50 uc NdNiO$_3$ film under the same conditions, which displays bulk-like properties. All samples are annealed at 600°C in flowing O$_2$ for 6 hrs following growth to ensure full oxidation of the Ni$^{3+}$ in the sample.

Figure 1(b-d) shows post-growth reflection high-energy electron diffraction (RHEED) images of the samples. Qualitatively, one can observe sharp RHEED spots with narrow vertical streaks, indicating highly crystalline samples with atomically flat surfaces. We also measure x-ray diffraction (XRD) and x-ray reflectivity (XRR) in a specular geometry using a lab-based Cu rotating anode source; the results are shown in Fig. 1(a). The XRD shows sharp superlattice peaks, attesting to abrupt interfaces, and displays no indication of additional phases (e.g. NiO). The XRR is fit using the GenX x-ray refinement program [36], giving values of root mean square interdiffusion and interfacial roughnesss of < ½ uc for all superlattices. This sample characterization provides evidence that the influence of extrinsic effects (such as point defects, intermixing and interfacial roughness) on the observed electronic and magnetic properties, reported below, is likely quite small.

At room temperature, the pseudocubic lattice constants for NdNiO$_3$, NdAlO$_3$ and LaAlO$_3$ are $a_{NdNiO_3}$ = 3.81 Å, $a_{NdAlO_3}$ = 3.74 Å and $a_{LaAlO_3}$ = 3.79 Å [37]. Thus, the average in-plane strain imposed by the substrate on the entire superlattice stack due to the lattice mismatch ranges from $\varepsilon$ = 0.40 to 0.96% (tensile) for the samples studied here. For this study, the important factor is the strain on the NdNiO$_3$ layer, which is relatively small: $\varepsilon$ = -0.52% (compressive).

Our first principles calculations use density functional theory (DFT) [38,39] with ultrasoft pseudopotentials [40-42] as implemented in the QUANTUM ESPRESSO software [43] to determine ground state structural and electronic properties of NdNiO$_3$ and (NdNiO$_3$)$_m$/(NdAlO$_3$)$_4$ superlattices [44]. The local density approximation (LDA) [45] is used to approximate the effects of exchange and correlation; comparison tests with the generalized gradient approximation (GGA) [46,47] show negligible differences, so results using LDA are reported below. We employ *k*-point sampling of the Brillouin zone equivalent to a 12×12×12 mesh for a 5 atom pseudocubic bulk unit cell. Band occupations are Gaussian broadened using a width of 0.03 eV. The plane wave cutoff is 35 Ry for wave functions and 280 Ry for the electron density. Structural relaxations are terminated when all components of atomic forces are < 0.03 eV/Å in magnitude. The simulated systems are periodic in all directions and biaxially strained to the theoretically computed pseudocubic lattice parameter of LaAlO$_3$ (3.71 Å). We consider superlattices with both 1×1 and *c*(2×2) interfacial unit cells, and the two only show small quantitative differences. In Sec. IV, we cite results using the *c*(2×2) unit cell, including octahedral rotations.

## III. Experimental Results

### A. Transport

The DC resistivity of the NdNiO$_3$ film and superlattices is acquired as a function of temperature (2-400K) using a Quantum Design PPMS Dynacool. The measurements are performed in a conventional Van der Pauw geometry with Au contacts sputtered on the corners of the sample.

Figure 2(a) displays the resistivity vs. temperature ($\rho$ vs. $T$) behavior of the samples. The 50 unit cell NdNiO$_3$ film undergoes a sharp metal-insulator transition with a transition temperature $T_{MI}$ = 156K, as determined by the minimum in the resistivity upon heating, and has a temperature hysteresis indicative of a first-order phase transition. The transition resembles the coincident metal-insulator and paramagnetic-antiferromagnetic phase transition of bulk NdNiO$_3$, but with a renormalized reduced transition temperature ($T_{MI,bulk}$ = 201K [10]) due to the clamping effect of the substrate, as observed in other studies [9,48,49]. For the NdNiO$_3$ films confined in superlattices, we find that a transition between a metallic ($d\rho/dT > 0$) and insulating ($d\rho/dT < 0$) region exists at $T_{MI}$ = 287K for *m*=4, albeit with a broad transition region and no discernable hysteresis. The *m*=2 and *m*=1 superlattices remain insulating at all measured temperatures.

To understand the transport behavior, we consider several models to describe the insulating regime in the *m*=1 superlattice. In particular, we consider (i) conventional semiconducting activated behavior, described by

$$\rho(T) \propto e^{E_g/k_B T}, \tag{1}$$

with $k_B$ = Boltzmann's constant and $E_g$ = the activation energy; (ii) 2D variable range hopping conduction, which follows

$$\rho(T) \propto e^{(T_0/T)^{\frac{1}{3}}}, \tag{2}$$

where $T_0$ is a characteristic localization temperature; and, (iii) the (non-adiabatic) small polaron hopping model, within which the resistivity behaves as

$$\rho(T) \propto T^{3/2} e^{W/k_B T}, \tag{3}$$

with $W$ representing the polaron binding energy [50].

The small polaron hopping model, Eq. (3), fits the data for $m=1$ well over the entire measured temperature range with $W$ = 114 meV, while deviations from the behaviors predicted from Eqs. (1) and (2) are apparent (Fig. 2(b-d)). This result points to the strong coupling of electrons with the lattice in the nickelate system [15] which may be enhanced due to confinement (as judged by the larger measured transport gap of the 1 layer superlattice relative to previous results on bulk or thick films of NdNiO$_3$, which report $E_g \approx$ 20-40 meV [51]).

In many materials, electronic or magnetic phase transitions can be identified by anomalies exhibited in quantities related to the derivative of $\rho(T)$. Based on the agreement for the $m=1$ superlattice, the small polaron hopping model described in Eq. (3) motivates the inspection of the energy parameter

$$W(T) = k_B \frac{d \ln(\rho/T^{3/2})}{d(1/T)} \tag{4}$$

as a function of temperature for all of the samples studied here. As mentioned above, when the small polaron hopping model is relevant, $W$ is the polaron binding energy. Otherwise, $W$ represents an energy scale for transport in the system whose behavior allows us to identify a temperature at which some electronic transition (if any) occurs. Details about the phase changes associated with this temperature are discussed in the sections below.

Figure 2(e-h) shows plots of $W(T)$ for the NdNiO$_3$ superlattices and 50 uc film. As expected from the fit in Fig. 2(b), the behavior of $W(T)$ for the $m=1$ superlattice is flat over the entire measured range. The NdNiO$_3$ film shows a sharp spike in $W(T)$ at $T_W$ = 152K, coinciding closely with $T_{MI}$. Such a spike in this type of energy parameter can be interpreted as a critical threshold in a percolation model of the first-order phase transition [48], and has also been associated with the onset of the charge-ordered, antiferromagnetic phase in nickelate films [52]. A similar peak in $W(T)$, though broadened, is found for the $m=4$ superlattice at $T_W$ = 144K, far below $T_{MI}$ determined above from the minimum in the resistivity. For the $m=2$ superlattice, we still

observe a (wide) peak in $W(T)$ centered around $T_W \sim$ 200K, whereas no resistivity minimum ($T_{MI}$) < 400K could be identified from Fig. 2(a).

As discussed below, the spike temperature of $W$, $T_W$, roughly coincides with charge and magnetic ordering temperatures determined from spectroscopy and scattering measurements. Hence, the transport measurements are the first indication that, contrary to the bulk-like NdNiO$_3$ film, there is a separation in characteristic temperatures, $T_{MI}$ and $T_W$, for the 2 and 4 layer superlattices, while no electronic transitions can be observed in 1 layer superlattices. These results point toward a confinement-induced crossover in the ground state properties of NdNiO$_3$.

## *B. X-ray absorption*

We examined changes in the electronic and orbital structure of the NdNiO$_3$ samples using soft x-ray absorption spectroscopy (XAS) and x-ray linear dichroism (XLD) at the Ni $L$ edge. Measurements are performed at beamline U4B of the National Synchrotron Light Source (NSLS) with linearly polarized x-rays. The polarization of the x-rays with respect to the crystallographic axes of the samples is altered by rotating the sample *in situ* while keeping the angle between the x-ray wavevector and the surface normal fixed (see Appendix A). All measurements are carried out in total electron yield mode and normalized to the incident x-ray flux. The Ni $L$ edge energy was calibrated by simultaneous measurement of a NiS powder. The energy resolution of the measurement at the Ni $L$ edge was $\Delta E/E \approx 5 \times 10^{-4}$.

Figure 3 shows the polarization-averaged XAS data for the NdNiO$_3$ film and superlattices measured at $T$ = 300K and 77K. For each sample, the Ni $L_3$ edge displays a splitting into two primary peaks, which we label *A* (lower energy, ~853 eV) and *B* (higher energy, ~854 eV). This type of peak splitting is observed throughout the nickelate series and has been associated with the charge-transfer energy separating the O *2p* and Ni *3d* states near the Fermi level [53]. The measured spectra for the 50 uc NdNiO$_3$ film agree well with previous work on bulk and thin film NdNiO$_3$ [54,55]. The small pre-peak visible in some samples centered at ~850 eV is the La $M_4$ white line coming from the LaAlO$_3$ substrate. Considering first the spectra at $T$=300K (Fig. 3(a)), as the thickness of the NdNiO$_3$ layer is lowered a number of trends can be observed. Most evidently, the intensity of peak *A* strongly decreases relative to peak *B*; a subtler effect is the narrowing of peak *A* relative to peak *B*.

While the peaks substantially overlap at room temperature, we see a pronounced separation of the two peaks at lower temperatures ($T$ = 77K, Fig. 3(b)) for the thicker NdNiO$_3$ layers. This separation results from both an increase in the energy splitting of the two peaks as well as a narrowing of peak *A*. The origin and implication of these feature changes is discussed in detail in Sec. IV. We note that, at the $L_2$ edge, large changes in the spectra for the different samples cannot be observed within the resolution of the experiment due to differences in the multiplet structure [56], so we focus our analysis on the spectral features at the $L_3$ edge.

To quantify these changes, we fit the Ni $L_3$ spectra with the sum of two peaks (*A* and *B*), each described by a pseudo-Voigt function with amplitude, *I*, full width at half maximum (FWHM), $\gamma$, center energy, $E_0$, and fixed Lorentzian-Gaussian mixing parameter, $\eta$. The parameters *I*, $E_0$, and $\gamma$ are varied for each peak with fixed $\eta$ to fit the $L_3$ edge region (851.5 to 854 eV). More information on the peak fitting can be found in Appendix A. The key quantities that we consider in the following are the peak height ratio, $Q_I = I_B/I_A$, the peak splitting, $\Delta E_0 = E_0^B - E_0^A$, and the peak widths, $\gamma_A$ and $\gamma_B$. These quantities represent the main features of the $L_3$ peak and are insensitive to the precise energy and intensity calibrations that are chosen. That is, one can reliably compare these quantities between samples and temperatures without requiring fine-tuned normalization or energy shifting.

We measure XAS at the Ni *L* edge for each sample as a function of temperature from *T* = 300K to 77K. The extracted values from the aforementioned two-peak analysis for these spectra are presented in Fig. 4. First, at *T*=300K, we notice an increase in the $L_3$ peak height ratio ($Q_I$) and a decrease in the low energy peak width $\gamma_A$ as the thickness of the NdNiO$_3$ layers is decreased, in agreement with our qualitative observations from Fig. 3. Considering $Q_I$ as a function of temperature, the NdNiO$_3$ film shows a clear decrease between 170 and 120K, while the *m*=1 and 2 superlattices exhibit no systematic change within the error bars. The *m*=4 superlattice, on the other hand, experiences a steady decrease in $Q_I$ with decreasing temperature for *T* < 220K. The same qualitative behavior is also observed in the temperature dependence of $\gamma_A$ (the high energy peak width, $\gamma_B$, remains temperature independent within experimental error). Interestingly, while the scale for $Q_I$ appears to be thickness dependent, $\gamma_A$ approaches the same value for all samples at low temperatures: $\gamma_A \approx 0.9 \pm 0.05$ eV. Hence, this may represent the intrinsic core-hole lifetime broadening in the insulating state. In addition, the onset of changes in $Q_I$ and $\gamma_A$ for the NdNiO$_3$ film and *m*=4 superlattice correspond to their respective metal-insulator transition temperatures, $T_{MI}$, observed from transport. Thus, we posit that the spectroscopic origins of the $Q_I$ and $\gamma_A$ temperature dependence may derive from the reduced screening and localization of charge carriers in the insulating state.

The peak splitting ($\Delta E_0$), on the other hand, exhibits different characteristics as a function of temperature. At *T* = 300K, $\Delta E_0$ is similar for all NdNiO$_3$ thicknesses with a value close to $\Delta E_0 \approx 1.5 \pm 0.05$ eV. As the temperature is reduced, $\Delta E_0$ remains relatively constant, aside from a distinct jump between *T*=170K and 120K for all samples except *m*=1. The magnitudes of the jumps are 0.23 eV, 0.10 eV, and 0.07 eV for the NdNiO$_3$ film, two-layer, and four-layer superlattices, respectively (corresponding to 15%, 7% and 4% of $\Delta E_0$ at *T*=170K). No such jump occurs for the *m*=1 superlattice. As discussed further in Sec. IV, a jump in $\Delta E_0$ is also found across the bulk *R*NiO$_3$ series upon entering the charge-ordered phase [53,57]. The temperature of the jump corresponds with the temperature, $T_W$, of the peak in the transport energy scale (see Fig. 3), which, as mentioned earlier, has also been connected to the onset of the ordered phase in NdNiO$_3$. Hence, the XAS measurements suggest that the charge-ordering temperature ($T_{CO}$) matches $T_W$ and is separated from the onset of the insulating region, $T_{MI}$, in the 2 and 4 layer superlattices. In Sec. IV, potential origins of the phase separation and the disordered insulating state are considered.

## C. Resonant x-ray scattering

Transport and XAS data point to changes in the electronic ordering as the thickness of the NdNiO$_3$ layer is reduced in the superlattices. In bulk NdNiO$_3$, a non-collinear (spiral) spin ordering of the Ni moments develops at temperature $T_N = T_{MI}$ [58,59] with wave vector $k_{AF}$ = (¼,¼,¼) in pseudocubic reciprocal lattice units (R.L.U.). A similar type of unusual antiferromagnetic (AFM) order is also observed in the insulating state of LaNiO$_3$/LaAlO$_3$ superlattices [24]. Here, we look for this type of magnetic structure using resonant soft x-ray scattering (RSXS) at the Ni $L$ edge, where a strong resonant enhancement of the magnetic scattering is expected [59].

RSXS measurements were carried out at beamline 4.0.2 at the Advanced Light Source using the measurement geometry depicted in Fig. 5(a) with a CCD detector. The energy resolution of the measurement is $\Delta E/E \approx 2 \times 10^{-4}$ at the Ni $L$ edge. The linear polarization of the incident radiation is rotated using an elliptically polarizing undulator without moving the sample. In all of the data presented, the scattering intensities are corrected for absorption by subtracting the fluorescent background in each CCD image.

Figure 5(b) shows the energy dependence around the Ni $L$ edge of the scattering intensity at momentum transfer $q$ = (¼,¼,¼) for the bulk-like NdNiO$_3$ film at $T$ = 30K for both $\sigma$ and $\pi$ incident polarizations (labeled $I_\sigma$ and $I_\pi$). One can observe a large resonant enhancement, particularly at the low energy peak of the Ni $L_3$ edge ($E \approx 853$ eV, corresponding to peak $A$ from the XAS analysis). The spectra agree with those observed in previous RSXS studies on NdNiO$_3$, which attribute the scattering to magnetic ordering [59]. In addition, we see that $I_\pi > I_\sigma$, which can be explained by the fact that the $\sigma_{in} \to \sigma_{out}$ scattering channel is forbidden for magnetic scattering. That is, $I_\pi = I_{\pi\pi} + I_{\pi\sigma}$, while $I_\sigma = I_{\sigma\pi}$, where $I_{\epsilon\epsilon'}$ indicates the scattered intensity with incident and scattered photon polarizations $\epsilon$ and $\epsilon'$. Although the exact ratio $I_\pi/I_\sigma$ depends on the azimuthal angle ($\psi$), which was not precisely measured here, the non-collinear magnetic ordering in compressively-strained NdNiO$_3$ gives $I_\pi/I_\sigma > 1$ at all $\psi$, in agreement with our measurements [24].

For the superlattices, we look for magnetic scattering along the $q$ = (¼,¼,L) rod, as we expect the $z$-axis ordering pattern to be modified relative to the bulk. Angular scans through the magnetic Bragg peak for each of the samples are shown in Figure 5(c). The data are taken at the Ni $L_3$ peak energy at $T$ = 30 K, and the peak with $L$ = ¼ is shown to facilitate comparison with the thick film. The resonant scattering signal is visible in the NdNiO$_3$ film and $m$=2 and 4 superlattices but is absent in the $m$=1 superlattice, indicating that the Ni sublattice loses its characteristic magnetic structure in the single NdNiO$_3$ layer superlattice. The energy profiles, including the polarization dependences, of the scattered intensity for the $m$=2 and 4 superlattices are essentially identical to that of the NdNiO$_3$ film, verifying an analogous magnetic origin of the scattering (see Appendix B). We note, however, that we cannot decipher the spin alignment along the (111) direction without mapping the azimuthal dependence of the scattered intensity. As has been demonstrated for confined LaNiO$_3$ layers, we speculate that a bulk-like non-collinear spin structure may persist in the confined ($m=2$ and 4) NdNiO$_3$ layers. In Sec IV, we present a model explaining the suppression of magnetic order in the $m=1$ structure, but also discuss other possible causes of the null result.

The integrated intensity ($I_{int}$) of the scattering decreases for decreasing NdNiO$_3$ layer thickness. At $T$ = 30K, $I_{int}$ for the $m=2$ superlattice is ~40% of that for the $m=4$ superlattice and ~4% of that of the 50 uc thick film. $I_{int}$ does not scale with the total number of NdNiO$_3$ layers in the samples, nor is the scaling compatible with a finite probing depth. Rather, $I_{int}$ appears to scale with $m$, the thickness of the NdNiO$_3$ layer in each repeat unit of the superlattice. Since the intensity of magnetic scattering $I_M \propto M^2$, where $M$ is the expectation value of each Ni magnetic moment, this finding suggests that the magnitude of the Ni moments is reduced for thinner NdNiO$_3$ layers.

We also observe differences in the peak width, which signals variation in the spin correlation length. By constructing a reciprocal space map (RSM) around the $q$ = (¼,¼,L) reflection, we can extract peak widths along the different reciprocal lattice directions. The in-plane spin correlation length is given by $\xi_S^{\parallel} = a/\gamma_S^{\parallel}$, where $\gamma_S^{\parallel}$ = FWHM of peak along $q_{x,y}$ in R.L.U. and $a$ = in-plane lattice constant. For the film, $\xi_S^{\parallel} \approx 20$ nm and decreases to 10-12 nm for the $m$ =2 and 4 superlattices. In all cases, the values for $\xi_S^{\parallel} \geq 25$ in-plane uc, illustrating the long-range nature of the in-plane AF order. For the superlattices, the RSM reveals that the scattering is extended in $q_z$ forming a rod of intensity characteristic of quasi-2D spin correlations (see Appendix B). The out-of-plane spin correlation length for the superlattices, $\xi_S^{\perp}$, determined from the width of the magnetic reflection in $q_z$, cannot be precisely ascertained because the measurement does not encompass the entire peak. We are, however, able to establish an upper bound of $\xi_S^{\perp} \leq 1.5$ nm (≈4 uc) for the $m$ = 4 superlattice, which indicates that the spin correlations are restricted to each separate NdNiO$_3$ block of the superlattice. In bulk NdNiO$_3$, it was determined that Nd magnetic moments are induced by the Ni moments [58], so it is a reasonable finding that interlayer magnetic coupling is absent through the NdAlO$_3$ spacer layer, which would likely require coupling through the Nd spins. This notion is corroborated by measurements at the Nd $M_{5,4}$ edge, which show no $q$ = (¼ ¼ ¼) diffraction signal.

To compare the magnetic ordering with the temperature scales observed in transport and XAS, we monitor the resonant $q$ = (¼,¼,L) reflection, at $L$ = ¼, while heating each sample from $T$ = 30-200K. As shown in Fig. 6, the integrated scattering intensity decreases monotonically with increasing temperature up to the transition temperature for all magnetically ordered samples. As noted above, the magnetic scattering intensity $I_m \propto m^2$, so we fit the integrated intensity to model the temperature evolution of the magnetic order parameter. For the 50 uc film, $I_{int}(T)$ can be fit well for $T<T_N$ by the mean field solution for a spin-½ classical antiferromagnet:

$$m(T) = m_0 B_S\left(\frac{3S}{S+1}\frac{T_N}{T}\frac{m}{m_0}\right), \tag{5}$$

where $B_S(x)$ is the Brillouin function, defined by

$$B_S(x) = \frac{2S+1}{2S} \coth\left(\frac{2S+1}{2S}x\right) - \frac{1}{2S}\coth\left(\frac{1}{2S}x\right). \tag{6}$$

For the fit, we assume $S = \frac{1}{2}$ and allow $T_N$ and $m_0$ to vary. We find $T_N \approx 138$K, slightly lower than $T_{MI}$. The superlattices show markedly different temperature dependences, which diverge strongly from the mean field solution. The superlattice integrated intensities follow a power law as a function of the reduced temperature, with

$$m(T) = m_0 \left(\frac{T_N - T}{T_N}\right)^\beta, \tag{7}$$

where $\beta \approx 0.5$, $T_N \approx 152$K ($< T_{MI}$) for the $m=4$ superlattice and $\beta \approx 1$, $T_N \approx 137$K for the $m=2$ superlattice. Such extended power law behavior beyond the critical region around $T_N$ has been observed previously and may originate from the enhancement of spin fluctuations in highly anisotropic 2D systems [60]. These results provide further evidence for the conclusion that the spin ordering in the superlattices is quasi-2D in nature.

The range of onset temperatures for the antiferromagnetism ($T_N \approx 130$-$150$K) matches the temperature associated with spectroscopic and transport anomalies, which supports the notion that these features are related to a concurrent charge and spin ordering transition. The data also corroborate the notion of the emergence of two distinct characteristic temperatures ($T_N = T_{CO} = T_W$ and $T_{MI}$) in confined $NdNiO_3$ layers in superlattices and the complete suppression of transitions in the single-unit cell thick layer. Possible origins of the separation of the metal-insulator transition from the charge/spin ordering transitions are discussed in Sec. IV.

### D. X-ray linear dichroism

A clue to the origin of the suppression of phase transitions in the $m=1$ superlattice can be found by probing the orbital symmetry using XLD. In this measurement, the polarization of the incident x-rays is varied to point along ($\vec{E} \perp \vec{c}$) or with a component normal to the sample surface ($\vec{E} \parallel \vec{c}$); the respective absorption intensities are labeled $I_{xy}$ and $I_z$. In the case of $Ni^{3+}$ in $NdNiO_3$, the $t_{2g}$ levels are full, so only the $e_g$ levels are probed at the Ni L edge. Then, $I_{xy}$ selectively measures transitions corresponding to Ni $2p \to 3d(x^2-y^2)$ states, and $I_z$ measures transitions from $2p \to 3d(3z^2-r^2)$ states. We use the difference between $I_{xy}$ and $I_z$ integrated over the Ni L edge to determine the relative occupations of the two Ni $e_g$ orbitals. $I_{xy} > I_z$ indicates a lower occupation of the $d(x^2-y^2)$ orbital, and hence an $e_g$ splitting with $d(3z^2-r^2)$ lower in energy than $d(x^2-y^2)$ (and vice versa when $I_{xy} < I_z$).

Figure 7(a,b) compares $I_{xy}$ and $I_z$ for the $m=4$ and 1 superlattices. As can be seen, $I_{xy} = I_z$ for the $m=4$ superlattice (as well as for the $m=2$ superlattice and the 50 uc film, which are not shown), but for the $m=1$ superlattice a dichroism emerges with $I_{xy} > I_z$. We track the dichroism by determining the peak height ratio dichroism, $\Delta Q_I = Q_{I_z} - Q_{I_{xy}}$, for each sample as a function of temperature. A positive value of this quantity suggests that the integrated

$I_{xy} > I_z$, and hence $E(d_{3z^2-r^2}) < E(d_{x^2-y^2})$. We see that for the *m*=1 superlattice, $\Delta Q_I$ is positive and largely constant with temperature, while for the 50 uc film and the *m*=2 and 4 superlattices, $\Delta Q_I \approx 0$ across the whole temperature range. This result establishes that there is a significant orbital polarization in the *m*=1 superlattice – with a lowering of the *d(3z²-r²)* energy level relative to *d(x²-y²)* – that is absent in samples with thicker NdNiO₃ layers. We note that the observed linear dichroism is likely not magnetic in origin due to the independence of $\Delta Q_I$ with temperature.

The existence of orbital polarization may explain the suppression of phase transitions and long range order in the *m*=1 superlattice. As will be discussed in Sec. IV, the connection between orbital polarization and suppression of long-range order supports a model of the phase transitions in the NdNiO₃ and thicker superlattices driven by a site-ordering instability.

## IV. Discussion

### *A. Theoretical calculations*

The experimental data presented above demonstrate a confinement-induced transformation in the ground state of NdNiO₃ from a charge and spin ordered phase to one lacking coherent long-range order. Furthermore, apparent orbital polarization is detected only in the *m*=1 superlattice, for which signatures of long-range charge or magnetic ordering are found to vanish. Two essential questions emerge based on these experimental observations: How is the electronic structure influenced by the NdNiO₃/NdAlO₃ heterointerface? And, what drives the suppression and separation of long-range ordered phases in the superlattices?  To help answer these questions, we turn to first-principles theory in conjunction with model Hamiltonian calculations of electronic spectra.

Table I summarizes the key electronic and structural parameters determined from relaxed DFT calculations of bulk NdNiO₃, the (NdNiO₃)₂/(NdAlO₃)₄ superlattice (*m*=2), and the (NdNiO₃)₁/(NdAlO₃)₄ superlattice (*m*=1). In agreement with the XLD (Sec. IIID), the *m*=1 superlattice shows a sizable negative Ni $e_g$ orbital energy splitting, $\Delta_{e_g} = E(d_{3z^2-r^2}) - E(d_{x^2-y^2})$, which is larger in magnitude than that of the bulk or the *m*=2 superlattice. Contrary to simple expectations, the orbital energy splitting is not correlated with changes in the local octahedral structure, as the apical and in-plane Ni-O bond lengths are identical for all the structures studied (see Table I). Instead, we find that $\Delta_{e_g}$ in the NdNiO₃/NdAlO₃ superlattices is due to an electrostatic effect: the increased positive ionic charge of the Al cation relative to Ni (Fig. 8). Even though Al and Ni both have the same formal 3+ charge state, Al is much less electronegative and Al-O bonds are more ionic than Ni-O bonds, both of which lead to the Al site being more positively charged.  A measure of this is to compare the occupancies of the apical O $2p_\sigma$ orbitals that point towards (and bond with) the cations. As Table I shows, the O $2p_\sigma$ orbitals that have more Al neighbors are significantly more occupied, demonstrating the more ionic nature of the Al-O bond. The increased positive charge on the Al creates a potential well for electrons in its vicinity. Since the interfacial Ni $d_{3z^2-r^2}$ electrons extend towards the Al and sample the region of lower potential, while the Ni $d_{x^2-y^2}$ do not tend to do so, we should find a lowering of the energy of the $d_{3z^2-r^2}$ orbital. The effect is strongest in the case of a NdAlO$_3$ interface on both sides of the Ni (*i.e. m=1*).

The difference in ionicity between Al-O and Ni-O bonds was noted in prior work [61-64], and the nature of the insulating *B*-site cation (Al in our case) is correlated with the Ni orbital polarization [63]. However, prior understanding of the effect of the ionicity centered solely on the reduced hole density and increased Madelung potential of the apical oxygen. We note that taking into account the behavior of the the apical oxygen alone would in fact predict the opposite direction of Ni orbital energy splitting to that which we observe, since an apical oxygen that is more electron rich and more negatively charged (reduced hole density) repels electrons and pushes up the energy of the Ni $d_{3z^2-r^2}$ orbital that points directly towards it. In contrast, the mechanism we propose is based on the direct electrostatic effect of the Al cation.. The positively charged Al lowers the electronic potential at nearby positions. The nearby apical oxygens become more negatively charged in an effort to dielectrically screen the field coming from the Al, but due to the finite and positive dielectric response, these screening effects only reduce the magnitude of the field and not its sign. Hence, the residual screened field from the Al provides net lowering of the electronic potential for the Ni $d_{3z^2-r^2}$ orbital. Given the generic nature of this new ionicity effect, one can imagine that it may be exploited at other materials interfaces to control relative orbital energies.

To gain further insight into the spectroscopic changes due to the interface, we carry out calculations of Ni *L* edge XAS using the charge-transfer multiplet model, which takes into account the crystal field and Ni-O hybridization (see Appendix C for details). The model can qualitatively reproduce the experimentally measured XAS for the *m*=1 superlattice and the 50 uc NdNiO$_3$ film (Fig. 13). From the extracted model parameters for each system, we find two important changes in the electronic structure. The magnitude of the $e_g$ energy splitting increases from $\Delta_{e_g} < 0.1$ eV (50 uc film) to $\Delta_{e_g} = -0.3$ eV (*m*=1), consonant with both the measured XLD and the DFT calculations. The Ni-O charge-transfer energy also increases from $\Delta_{pd} = 1.4$ eV (50 uc film) to $\Delta_{pd} = 2.2$ eV (*m*=1). The charge-transfer energy for Ni$^{3+}$ is defined as

$$\Delta_{pd} = E(d^8\underline{L}) - E(d^7), \tag{8}$$

where $\underline{L}$ is a ligand (in this oxygen) hole, giving the energy cost for transferring an electron from the O *p* to the Ni *d* states. The increase in $\Delta_{pd}$ for the *m*=1 superlattice signals a less covalent Ni-O bond. Again, the DFT calculations reveal an anisotropic Ni-O hybridization, which is evident in the increased apical O 2*p* occupancy for the *m*=1 superlattice (Table I) and an increase in the O 2*p* on-site energy by ~0.5 eV, in general agreement with the multiplet calculations. The origin of the diminished Ni-O hybridization is the confinement of electrons to the Ni-O-Ni plane, as hopping along the Ni-O-Al direction into the closed shell Al$^{3+}$ ion with high energy unoccupied states is suppressed (Fig. 8). Similar interfacial reconstructions of the Ni-O states have been found in LaNiO$_3$-based superlattices [61,64-66].

The interfacial coupling effects mentioned above are most significant for the $m$=1 superlattice, where all apical oxygens comprise Ni-O-Al bonds. The combination of large $e_g$ orbital splitting and anisotropic hybridization leads to the orbital polarization observed in Sec. IIID. As discussed below, the orbital polarization and reduced covalency explain the absence of long-range order in the $m$=1 superlattice, and instead stabilize a charge-transfer or Mott insulating ground state. The interfacial layers of the $m$=2 and 4 superlattices will also be affected by the interfacial coupling; however, when $m$>1, at least one apical Ni-O bond includes a "bulk-like" O atom within the NdNiO$_3$ layer (*i.e.* with Ni-O-Ni rather than Ni-O-Al bonds), which decreases the $e_g$ energy splitting, increases Ni-O hybridization, and reduces the average orbital polarization (see Table I and Fig. 7). As a result, the electronic states are more bulk-like, but reduced dimensionality causes the observed unusual phase behavior, which we discuss further below.

## *B. Phase diagram and qualitative model*

Having identified and understood the aforementioned changes in electronic structure in the NdNiO$_3$/NdAlO$_3$ superlattices, we seek to link them to the observed phase behavior. It has also been found in other confined nickelate systems that collective ordering behavior is greatly altered relative to the bulk. For instance, in the case of LaNiO$_3$ (which is a paramagnetic metal in bulk) a metal-insulator transition emerges accompanied by antiferromagnetic ordering in superlattices with ultrathin (2 uc) nickelate layers. Theoretical calculations [19] and photoemission measurements [67] also suggest the presence of antiferromagnetic order in the ground state of monolayer LaNiO$_3$. More recently, Raman spectroscopy performed on PrNiO$_3$ – which exhibits electronic and magnetic behavior in the bulk akin to NdNiO$_3$ but with a lower $T_{MI}$ – demonstrates the loss of charge order in compressively strained superlattices but preservation of the spin ordering transition occurring at the minimum in the resistivity [25]. A temperature-independent $e_g$ orbital polarization with enhanced Ni $d(3z^2-r^2)$ occupation is also found in these superlattices [32]. Together, these observations have been interpreted as evidence in support of a spin-density wave in the low-temperature phase arising from Fermi surface nesting that is enhanced by spatial confinement, as proposed in Refs. [16,17]. It should be noted that the ground state in $R$NiO$_3$ films and superlattices depends on strain. The metal-insulator transition in thin films can be shifted by epitaxial strain (in analogy to hydrostatic pressure in the bulk) [9]. In addition, it has been reported that the orbital polarization under compressive strain diminishes and gives way to a charge-disproportionated state under tensile strain [19,61,62,66]. The relationship between orbital polarization and charge ordering plays an important role in the physics of the nickelates [68] and will be discussed below and in the context of the observed phase behavior of our NdNiO$_3$ superlattices.

To tie the interfacial electronic reconstruction to the observed phase behavior, we develop a model for the ordered state. There has been substantial experimental evidence and theoretical support for the existence of two inequivalent Ni sites in the insulating ground state of bulk

NdNiO$_3$ (and other $R$NiO$_3$ with smaller $R$), accompanied by ordering of the O or Ni charge. The O-centered model emphasizes the ordering of Ni-O hybridized $d^8\underline{L}$ states ($2d^8\underline{L} \rightarrow d^8 + d^8\underline{L}^2$), while the Ni-centered model considers the Ni $e_g$ states alone ($2e_g^1 \rightarrow e_g^2 + e_g^0$). Our XAS measurements average over all Ni sites, and the calculations above are based on a single Ni-O cluster; hence, we cannot observe this inequivalence directly. Recent XAS and resonant x-ray diffraction measurements on a NdNiO$_3$ film display Ni $L_3$ peak splitting in the insulating state without the associated Ni charge disproportionation (though O-centered ordering is not ruled out) [69]. Nonetheless, we find an enlargement of $\Delta E_0$ for the NdNiO$_3$ film and $m$=2 and 4 superlattices as a function of temperature (Fig. 4), overlapping with the spin ordering transition. Such an enlargement is observed in the bulk across $T_{MI}$ coincident with the charge-ordering transition, but no such equivalent spectral changes are observed at $T_N$ for those systems with $T_{MI} \neq T_N$. Further, the fact that the XAS transitions for the superlattices occur below $T_{MI}$ suggests the existence of an electronic reconstruction separate from the metal-insulator transition, which likely stems from Ni or O-type charge ordering. Thus, we hypothesize that the observed spin ordering transitions (Fig. 6) are accompanied by charge ordering, and neither ordering exists in the case of the $m$=1 superlattice. Determining the exact nature of the long-range order requires future structural or spectroscopic measurements, but a schematic picture is presented below.

Figure 9 summarizes the phase behavior of the NdNiO$_3$ superlattices studied here as a function of the number of NdNiO$_3$ layers. As in the case of LaNiO$_3$-based and PrNiO$_3$-based superlattices, our NdNiO$_3$ superlattices show an increase in the metal-insulator transition temperature with confinement. Separation of the spin ordering transition from the onset of the insulating state of up to has also previously been reported. However, in contrast to other nickelate systems, the confinement in NdNiO$_3$ leads to a very large separation of $T_{MI}$ from $T_{CO}$ (~150 K for $m$=4 and >200 K for $m$=2), which remains coincident with $T_N$ until the eventual collapse of the bulk-like charge/spin ordered state altogether for the single NdNiO$_3$ layer superlattice ($m$=1). This hidden phase of the fully 2D $m$=1 superlattice had not been characterized previously and may be present in other nickelates when confined to single layers.

Based on a model of charge and spin-ordering on inequivalent Ni sites, the suppression of phase transitions in the $m$=1 superlattice can be understood on the grounds of the orbital degeneracy breaking measured using XLD and understood theoretically above. The situation is schematically depicted in Fig. 10; for concreteness, we focus on the O charge ordering scenario below. In bulk NdNiO$_3$, each site's orbital degeneracy is preserved upon entering into the ordered phase (Fig. 10(a,b)). The $d^8$ and $d^8\underline{L}^2$ sites ($e_g^2$ and $e_g^0$ sites in the Ni-centered scenario) have spin $S$=1 and $S$=0, respectively, and the energy of the system will be lowered by an amount related to the Hund's coupling energy ($J$). However, if the $e_g$ orbital energy splitting ($\Delta_{e_g}$) is sufficiently large in comparison to $J$, a low-spin configuration ($S$=0) on the $d^8$ site can arise. Rather than forming an ordered lattice of inequivalent $S$=0 sites, it is more favorable for the charge ordering and the accompanying antiferromagnetic alignment to be suppressed for large $\Delta_{e_g}$ (Fig. 10(d)). The system would then remain in a mixture of $d^7$ and $d^8\underline{L}$ configurations, as in the high temperature disordered state [68]. This case likely occurs in the $m$=1 superlattice, and the observed increase in the charge-transfer energy ($\Delta_{pd}$) suggests the subsequent development of a conventional charge-transfer or Mott insulating state, potentially reinforced by enhanced electronic correlations in 2D. Alternatively, a phase separated state or a fluctuating charge-ordered state may be favored. These scenarios are consistent with the observed temperature independence of the XAS and RSXS for the *m*=1 superlattice, and further time or spatially resolved studies are needed to provide more insight. As mentioned above, this type of competition between orbital and charge order has been directly observed in strained LaNiO$_3$ films where the charge ordering (breathing mode) distortion is only active under tensile strain in the absence of $e_g$ orbital polarization.

An important condition inherent to this picture is that the energy gain due to orbital polarization is larger than that due to the charge disproportionation for the *m*=1 superlattice. A simple physical justification is provided here in the localized limit. Relative to the unordered phase with a single degenerate $e_g$ electron per Ni site, an orbitally polarized state occupied by a single electron would have an energy of $-\Delta_{e_g}$, where $\Delta_{e_g} \sim 0.3$ eV for the *m*=1 superlattice. The energy of the charge ordered state is estimated as $E_{CO} \sim U - 3J$ (ignoring differences in the on-site energy from the octahedral breathing distortion), where *U* is the on-site Coulomb energy [20]. In this picture, the tendency towards charge ordering in bulk nickelates (with $\Delta_{e_g} \sim 0$ eV) is explained by screening, which is enhanced by relatively small charge-transfer energy ($\Delta_{pd}$) and large $e_g$ bandwidths. The screening makes for a modest effective *U* and, hence, a small negative value for $E_{CO}$. That $E_{CO}$ is relatively small is supported by the fact that the charge ordered state has a small band gap of less than ~0.1 eV and low transition temperature of ~200 K for bulk NdNiO$_3$ [70]. As dimensionality is reduced to the 2D limit in the *m*=1 superlattice, the bandwidth reduction and increase in $\Delta_{pd}$ will reduce the screening of *U* and raise $E_{CO}$. Thus, comparing $E_{CO}$ and $\Delta_{e_g}$ under these conditions, it is likely that the charge ordered state could become unstable with respect to the orbitally polarized state.

Unlike the *m*=1 superlattice, the *m*=4 and 2 superlattices lack orbital polarization (Fig. 7), despite containing confined quasi-2D NdNiO$_3$ layers. In the case outlined above, the observed separation of $T_{MI}$ and $T_N/T_{CO}$ for these superlattices then results chiefly from dimensional confinement (Fig. 10(c)). The decrease of the average bandwidth and increase in the average charge-transfer energy due to confinement pushes the system towards the more insulating side of the phase diagram (akin to a smaller *R* ion). The measured high temperature insulating state may thus result from a bandwidth narrowing transition without charge or spin order, creating a charge-transfer gap between Ni *3d* and O *2p* bands [10]. Bulk-like order dominates at low temperatures, but large charge and spin fluctuations due to the reduced dimensionality can have the effect of broadening the metal-insulator transition [71], as observed in Fig. 2(a). This 2D fluctuation scenario is supported by the observation of non-saturating magnetic scattering for the superlattices, as noted in Fig. 6. Photoemission measurements on LaNiO$_3$ thin films also provide evidence for fluctuating order in proximity to a metal-insulator transition, as is the case for the *m*=4 and 2 superlattices [67]. We note that in the case of single layer LaNiO$_3$, it has been suggested that long-range order is stabilized in 2D [19,67]; however, this situation may be ruled out in our case based on the observations for the $m$=1 NdNiO$_3$ superlattice, possibly due to greater orbital polarization.

The model presented above offers one potential explanation of the behavior we observe in NdNiO$_3$/NdAlO$_3$ superlattices. The observed lack of charge and magnetic order in the $m$=1 superlattice may also be explained as a consequence of the 2D nature of the NdNiO$_3$ layers. In truly 2D layers with only nearest-neighbor interaction, one would expect that long-range ordered antiferromagnetism would not develop at finite temperature. Of course, many quasi-2D systems (*e.g.* cuprates) with residual interlayer couplings and/or next-nearest-neighbor interactions possess magnetic phase transitions. Alternatively, a reconstruction of the magnetic interactions in 2D could lead to antiferromagnetism with an ordering vector different from the bulk-like $q$ = (¼ ¼,$L$). It should also be noted that any residual structural defects in the superlattice, in conjunction with strong 2D fluctuations and a diminished local moment, might broaden the diffraction peak and push the signal below the noise floor of the experiment. Other sensitive magnetic probes, such as muon spin rotation or x-ray magnetic circular dichroism, would be valuable to help rule out the existence of antiferromagnetic correlations in the single layer NdNiO$_3$ heterostructure.

These results provide new and important insight into lively theoretical discussions regarding the nature of the confined nickelate ground state, which has not been extensively explored experimentally. Mean field approaches using a two-band ($e_g$ only) model as well as results including Ni-O charge transfer both predict an antiferromagnetic ground state with $q$=(¼,¼) ordering for single confined nickelate layers ($m$=1) over a range of interaction parameters. The observed phase suppression in our NdNiO$_3$/NdAlO$_3$ superlattices thus calls for further theoretical consideration of the single layer limit. For thicker nickelate layers, the $p$-$d$ model calculations qualitatively describe the observed antiferromagnetic ordering and the tendency towards insulating behavior with decreasing thickness. However, a paramagnetic to antiferromagnetic transition is only present in the calculations for $m$≥3, while similar magnetic phase transitions are measured here for $m$=2 and $m$=4. Hence, while many aspects of the nickelate superlattices can be understood based on simple low-energy models, ancillary effects, such as interfacial reconstructions, appear to be crucial ingredients to describe the experimental situation.

A final interesting result is that the charge/spin ordering temperature is nearly the same for all of the samples (except $m$=1, which displays no ordering). One potential explanation is that the ordering temperature could be primarily determined by the in-plane Ni-O bond properties (bond lengths and bond angles). As the in-plane epitaxial strain on the NdNiO$_3$ layers is the same for all the samples, the non-interfacial Ni-O structural properties will likely be similar. To more accurately assess this possibility, a detailed structural characterization of the superlattices will be a useful avenue for future study.

## V. Conclusions

Employing a variety of experimental techniques, we reveal the hidden electronic and magnetic phases of MBE-grown $NdNiO_3$/$NdAlO_3$ superlattices. From electrical resistivity measurements, we find an increase in $T_{MI}$ for confined $NdNiO_3$ layers, with 1 and 2 uc thick remaining insulating up to 400K. Above 1 uc, anomalies in transport and XAS associated with charge ordering at temperatures below $T_{MI}$ coincide with the onset of 2D antiferromagnetic order determined from RSXS. Hence, a charge and spin ordered ground state exists in superlattices down to 2 uc $NdNiO_3$, while no indications of such ordered phases exist in 1 uc $NdNiO_3$.

X-ray linear dichroism measurements help explain this behavior by revealing that only the 1 uc $NdNiO_3$ superlattice experiences a significant Ni $e_g$ orbital polarization with higher occupancy of $3d(3z^2-r^2)$ states. Furthermore, first-principles theory and cluster calculations reveal the existence of large $e_g$ orbital splitting reduced Ni-O hybridization at the interface. These changes inhibit the Ni site ordering, which accompanies the electronic and magnetic phase transitions in the bulk. Thus, charge and spin order is suppressed in 1 uc $NdNiO_3$ superlattices. And, unlike the bulk, the ordering temperatures in the thicker superlattices are separated from the metal-insulator onset temperature due to the phase competition imposed by the interface and the enhancement of fluctuations in reduced dimensions.

These experiments illustrate novel collective phenomena in strongly correlated materials as a result of dimensional confinement and interfacial coupling. The behavior differs from previous reports on nickelate superlattices, expressing the unique sensitivity of these confined systems. In addition, the results illuminate the importance of orbital degeneracy, dimensionality, and charge-disproportionation in the ground state of the nickelates. For future study, it may be interesting to investigate different strain states and layer compositions in an effort to unify the understanding of phase control in nickelates and other correlated systems through heterostructuring.

## Appendix

### *A. XAS measurement and data analysis*

In the experimental setup for the XAS and XLD measurements, the polarization of the incident x-ray ($\vec{E}$) is fixed, so the sample must be rotated *in situ* to determine linear dichroism. In order to maintain a constant incident angle (α) while rotating the crystallographic axes about the incident beam direction, the sample is placed on a Mo wedge cut at an angle of 20° with respect to the incident beam direction. In this way, the x-ray probing depth is kept constant for the absorption measurement as the angle ($\theta$) between $\vec{E}$ and the sample *a* (or *b*) axis is varied.

The measurement geometry is such that for one configuration ($\theta = 0°$), $\vec{E}$ lies entirely in plane, while in the rotated configuration ($\theta = 90°$), $\vec{E}$ has both in-plane and out-of-plane components. For the purposes of the XLD measurements, the in-plane absorption intensity is given by

$I_{xy} = I(\theta = 0°)$, while a geometrical correction is necessary to extract the out-of-plane absorption intensity $I_z \approx 1.132(I(\theta = 90°) - 0.117 I_{xy})$ [30,72].

The raw Ni $L$ edge XAS data are aligned in energy with respect to a simultaneously measured NiS standard. The Ni $L_3$ peak of the standard is calibrated to 853.2 eV [73]. The absorption intensity at each energy is normalized to the incident flux, measured upstream of the sample with a gold mesh. In addition, the background in the Ni $L$ edge region is subtracted by fitting the pre-edge region to a linear polynomial and fitting the steps due to continuum excitations at the $L_3$ and $L_2$ edges. The step edge background is modeled by a superposition of arctangent and error functions, with the $L_3/L_2$ step height ratio of $\sim 2/1$.

The analysis presented in Sec. IIIB is carried out by fitting the (background corrected) Ni $L_3$ edge to a double peak profile. Each peak is modeled with a pseudo-Voigt function, defined as

$$P(E; A, E_0, \gamma, \eta) = I[\eta L(E; E_0, \gamma) + (1 - \eta) G(E; E_0, \sigma)], \tag{9}$$

where $L(E; E_0, \gamma)$ and $G(E; E_0, \sigma)$ are Lorentzian and Gaussian profiles centered around $E_0$ with full widths at half maximum (FWHM) $\gamma$ and $\sigma = \frac{\gamma}{2\sqrt{2 \ln 2}}$, respectively. To constrain the number of fitting parameters, the value of the Lorentzian-Gaussian mixing parameter, $\eta$, is fixed by fitting the $L_3$ edge of a NiS reference to Eq. (5). The value $\eta = 0.23$ extracted from this reference fit is used for all Ni $L_3$ fits in this work. In some of the samples, a pre-peak at 850 eV can be noticed originating from the La $M_4$ absorption edge of the LaAlO$_3$ substrate. To account for this feature, we compare the extracted parameters from the two-peak fits to those found using three-peak fits, which include the La $M_4$ pre-peak, and we find negligible differences. Hence, the analysis in the main text is based on the two-peak model, which generally provides better fits to the experimental data. Examples of the two-peak fit to the Ni $L_3$ edge at low and high temperature for the 50 uc NdNiO$_3$ film and the *m*=1 superlattice are displayed in Fig. 10.

### B. Additional magnetic scattering data

Energy scans through the Ni $L$ edge of the *q* = (¼ ¼ ¼) magnetic scattering intensity are shown in Fig. 10(a,b) for the *m*=4 and 2 superlattices. The resonant enhancement, which is particularly pronounced at the lower energy $L_3$ edge, is similar to that found for the bulk-like 50 uc NdNiO$_3$ film (Fig. 5(b)). In both cases, $I_\pi/I_\sigma > 1$, which is consistent with non-collinear magnetic ordering, as has been found previously in NdNiO$_3$ and LaNiO$_3$/LaAlO$_3$ superlattices [24,59]. Note, however, that $I_\pi/I_\sigma$ differs for the two superlattices (and from the 50 uc film). The azimuthal angle ($\psi$) is set to be close to 45°; however, deviations of ±10° may lead to discrepancies in $I_\pi/I_\sigma$. Alternatively, the differences in $I_\pi/I_\sigma$ observed for the different samples may manifest differences in the relative spin alignments.

Reciprocal space maps (RSMs) are constructed from the CCD images obtained during angular scans along the (111) direction through the magnetic Bragg reflection. Figure 10(c,d) shows the

RSMs around $q = $ (¼ ¼ ¼) for the $m=4$ superlattice and the 50 uc NdNiO$_3$ film. The magnetic scattering intensity for the $m=4$ superlattice (Fig. 10(a)) extends in $q_z$ beyond the limits of the map. The extended peak width provides an upper bound to the out-of-plane spin correlation length, which indicates that the magnetic order is limited to a single superlattice layer, as discussed in Sec. IIIC. The scattering for the 50 uc film, on the other hand, shows a sharp central peak and additional finite thickness oscillations, revealing an out-of-plane spin correlation length $\xi_s^\perp \approx 15$ nm, which is nearly equal to the thickness of the film ($t_{film}$ = 19 nm).

## C. Charge transfer multiplet calculations

The results of Sec. IIIB, summarized in Fig. 4, demonstrate that the electronic ordering transitions manifest as changes in the spectral features in the Ni $L$ edge. To develop a detailed understanding of the electronic origin of these features, we perform analysis of the Ni $L$ edge XAS and XLD spectra using small cluster calculations within a charge-transfer multiplet framework to describe the covalent mixing between Ni 3$d$ and O 2$p$ states, as implemented in the XCLAIM program [74].

Following the procedure for calculating bulk $R$NiO$_3$ spectra in Ref. [53], we consider ground state cluster configurations of the form, $|\psi\rangle = \alpha|d^7\rangle + \beta|d^8\underline{L}\rangle + \gamma|d^9\underline{L}^2\rangle$, where $\underline{L}$ represents a ligand (in this case oxygen) hole and $\alpha^2 + \beta^2 + \gamma^2 = 1$. In all of the calculations, the Slater integrals for the electron-electron interaction are reduced to 80% of the Hartree-Fock values to account for intra-atomic screening. The isotropic parts of the core hole potential and valence Coulomb potential are set to $F_{cv}^0 = 7$ eV and $F_{vv}^0 = 6$ eV. The Slater-Koster terms parameterizing the Ni 3$d$ – O 2$p$ overlap are $(pd\sigma) = 1.3$ eV and $(pd\pi) = -0.515$ eV. A Gaussian broadening of 0.5 eV and Lorentzian broadenings of 0.5 eV for the $L_3$ edge and 1.0 eV for the $L_2$ edge are included to account for instrumental resolution and core-hole lifetimes. The remaining parameters – the charge transfer energy (Δ) and crystal field splittings (10$Dq$, $Ds$, $Dt$) – are adjusted to yield the best match to the experimental data. Since the $L_3$ peak height ratio ($Q_l$) shows the largest deviation between samples in the measured XAS, this was the primary quantity used to determine the suitability of the calculations. We also find that the trends in the peak widths ($\gamma_A$) and peak splittings ($\Delta E_0$) are reproduced in the calculated spectra.

The low temperature XAS data from the 50 uc thick NdNiO$_3$ film can be reproduced (Fig. 13) by considering an octahedral configuration ($O_h$ symmetry) with $e_g$-$t_{2g}$ ionic crystal field splitting of 10$Dq$ = 1.8 eV and a charge-transfer energy $\Delta_{pd} = 1.4$ eV. To account for the peak shape changes in the $m=1$ superlattice, a tetragonal configuration ($D_{4h}$ symmetry) must be used with $Ds$ = 0.075 eV and $Dt$ = 0 eV, which leads to an $e_g$ ionic crystal field splitting $\Delta_{e_g} = -0.3$ eV with $E(d_{3z^2-r^2}) < E(d_{x^2-y^2})$, and an increased charge-transfer energy $\Delta_{pd} = 2.2$ eV. Importantly, a non-zero $e_g$ splitting is necessary to generate an orbital polarization (as observed in experiment), but adding only an $e_g$ splitting while maintaining $\Delta_{pd} = 1.4$ eV cannot reproduce the experimentally observed lineshape. We remark that instead of changing $\Delta_{pd}$, uniformly reducing the $p$-$d$ overlap integrals also changes $Q_l$, in general agreement with experiment. A combination of both effects is most likely present, but as either case represents an effective reduction in Ni-O hybridization, we focus only on changes in $\Delta_{pd}$ in the discussion of Sec. IV for simplicity.

## Acknowledgments

Work at Yale supported by AFOSR and NSF under MRSEC DMR-1119826 (CRISP). J.L.H. and M.L.T. acknowledge support from the Office of Naval Research under contract number N00014-14-1-0058. Use of the Advanced Light Source is supported by the Director, Office of Science, Office of Basic Energy Sciences, of the U.S. Department of Energy under Contract No. DE-AC02-05CH11231. The National Synchrotron Light Source at Brookhaven National Lab was supported by the Office of Science, Office of Basic Energy Sciences, of the US Department of Energy under Contract No. DE-AC02-98CH10886. J.L.H. thanks Steven R. Spurgeon and William Ratcliff for helpful discussions on magnetic characterization.

## Figures

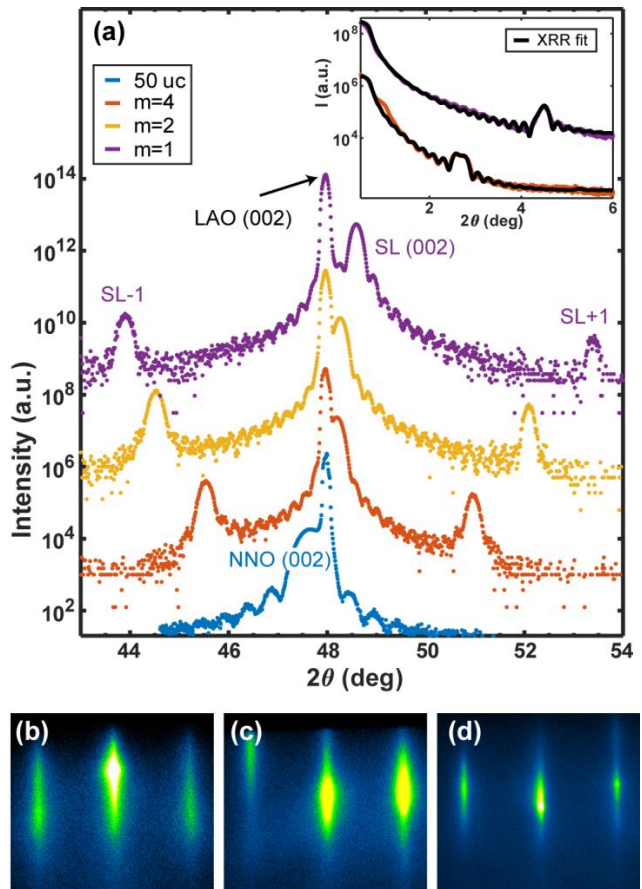

Fig. 1 (a) X-ray diffraction ($\vartheta$-$2\vartheta$) scan around the (002) Bragg peak for NdNiO$_3$ film (NNO) and superlattices (SL) grown on LaAlO$_3$ (LAO). Inset shows x-ray reflectivity for $m=4$ and $m=1$ superlattices with simulated fits. (b-d) Post-growth RHEED patterns for (b) $m=4$, (c) $m=2$ and (d)

*m*=1 superlattices.

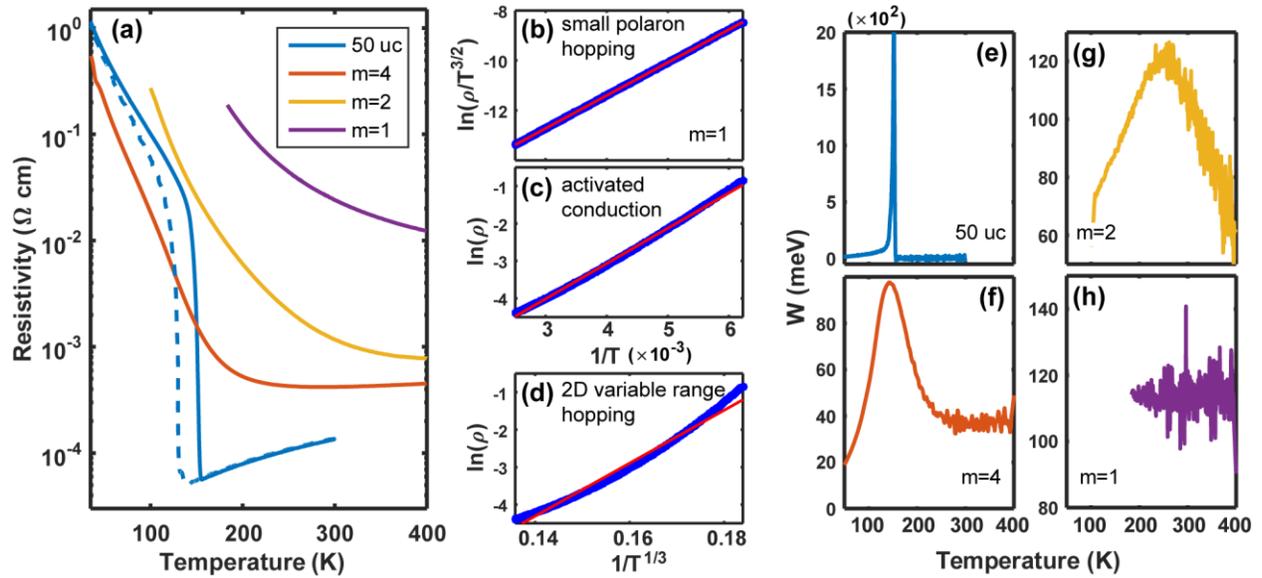

Fig. 2. (a) Resistivity vs. temperature for NdNiO$_3$ film and superlattices (solid and dashed line for 50 uc film shows resistance measured during heating and cooling, respectively). (b-d) Resistivity (*ρ*) vs. temperature for *m*=1 superlattice (blue circles) and linear fit (red line) to (b) small polaron hopping (Eq. 1), (c) activated conduction (Eq. 2) and (d) variable range hopping (Eq. 3) models. (e-h) Polaron energy, *W*, determined by Eq. 4, for the (e) 50 uc film and (f) *m*=4, (g) *m*=2, and (h) *m*=1 superlattices.

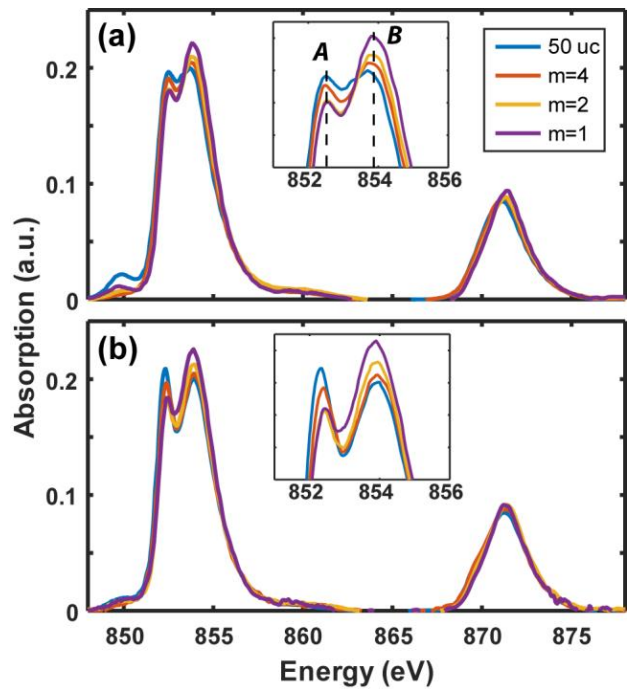

Fig. 3. (a) $T$ = 300 K and (b) $T$ = 77 K Ni $L$ edge XAS data. Insets provide a zoom-in of $L_3$ multiplet structure for respective temperatures, showing splitting of peaks A and B.

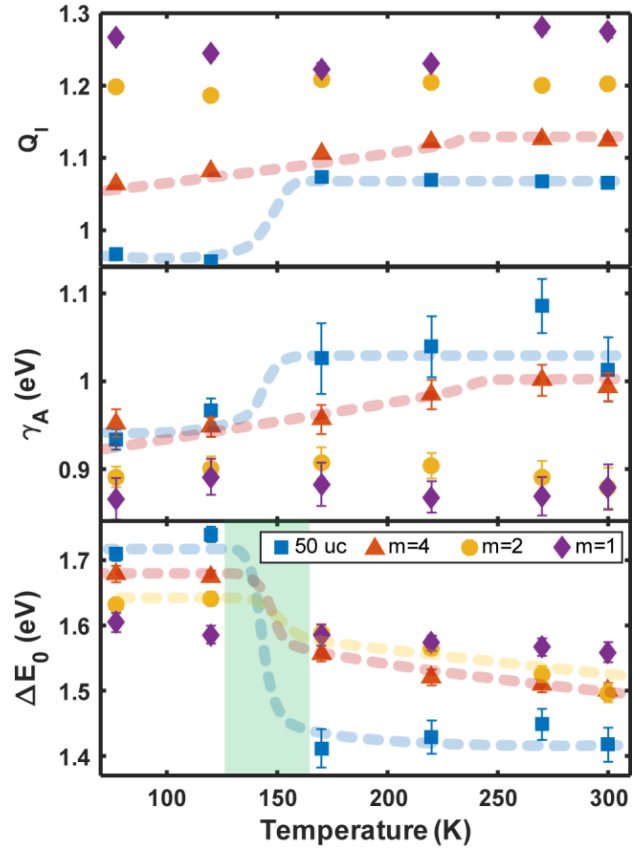

Fig. 4. Spectral parameters extracted from fits of Ni $L_3$ XAS data for NdNiO$_3$ film and superlattices as a function of temperature ($Q_I$ = peak height ratio, $\gamma_A$ = FWHM of peak A, $\Delta E_0$ = peak splitting). Error bars are determined from fit uncertainties. Dashed lines are guides to the eye, and green shading indicates the temperature region of the $\Delta E_0$ transition.

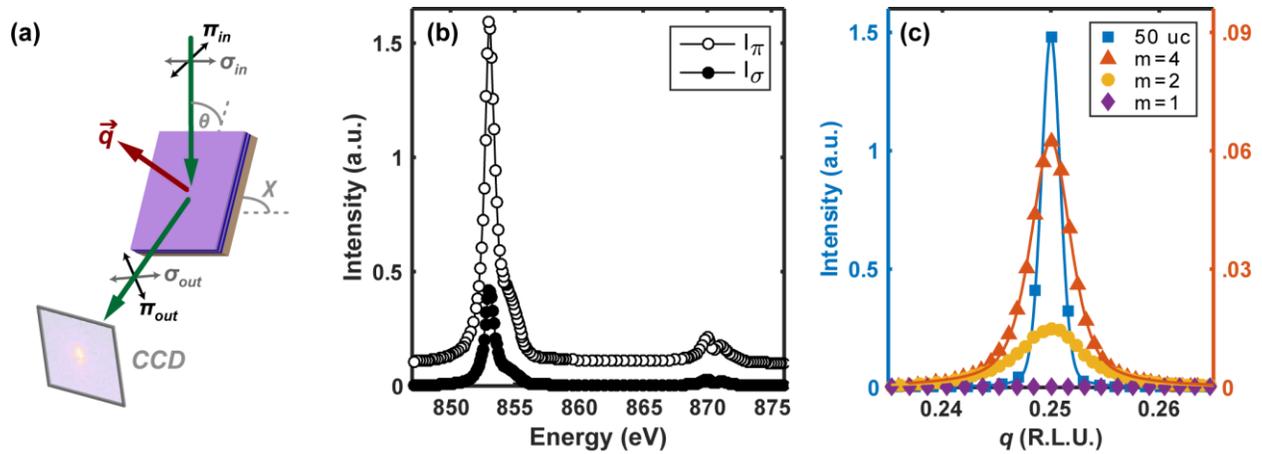

Fig. 5. (a) Schematic of RSXS measurement (fixed $\chi = 45°$, varied $\vartheta$). $\chi$ and $\vartheta$ are the rotation angles of the sample about the incident beam direction. (b) Energy dependence of $q = (¼,¼,¼)$ diffraction intensity through the Ni $L$ edge for the 50 uc NdNiO$_3$ film at $T$ = 30 K. (c) Angular scan along the (111) direction through the magnetic Bragg reflection at the Ni $L_3$ peak energy ($E$ = 853 eV) and $T$ = 30 K (symbols = normalized data and solid lines = Lorentzian fits).

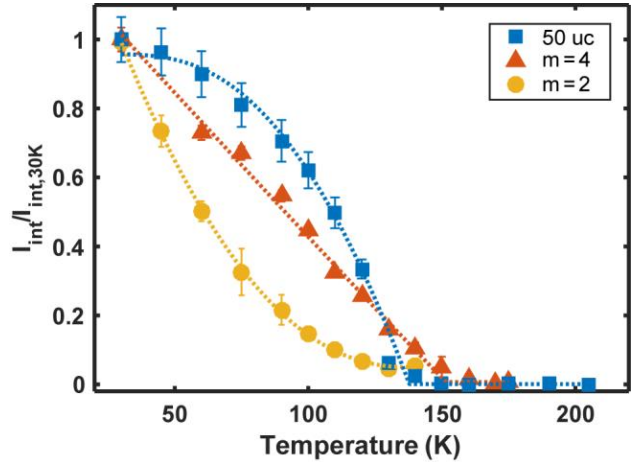

Fig. 6. Temperature dependence of the integrated intensity normalized at $T$ = 30 K for magnetically ordered NdNiO$_3$ film and superlattices. Dashed lines are fits to models of the magnetic order parameter (see text for details) with $T_N$ = 138 K, 152 K and 137 K for the 50 uc film, $m$=4 and $m$=2 superlattices, respectively.

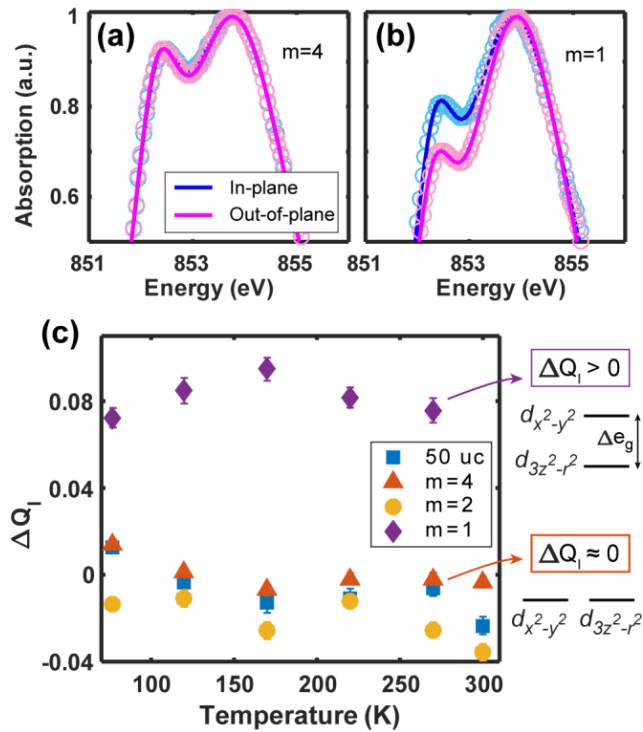

Fig. 7. (a) In-plane ($I_{xy}$, $\vec{E} \perp \vec{c}$) and (b) out-of-plane ($I_z$, $\vec{E} \parallel \vec{c}$) absorption spectra for $m=4$ and $m=1$ superlattices at $T = 270$ K zoomed in around the Ni $L_3$ peak. Measured data and two-peak fits are shown as open circles and solid lines, respectively. (c) Peak height ratio dichroism ($\Delta Q_I = Q_{I_z} - Q_{I_{xy}}$) as a function of temperature for NdNiO$_3$ film and superlattices. The right side of (c) shows schematic energy level diagrams corresponding to positive and zero $\Delta Q_I$.

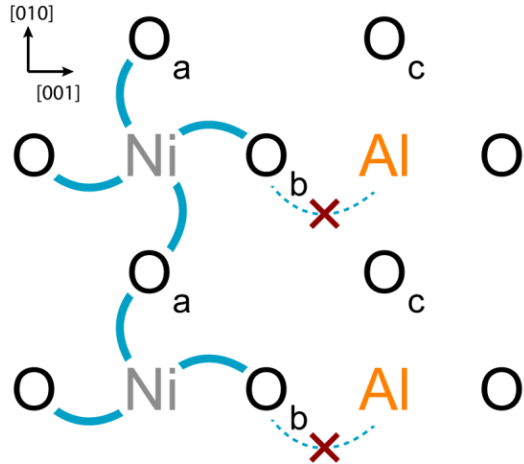

Fig. 8. Schematic of the NdNiO$_3$/NdAlO$_3$ interface depicting the metal-oxygen bonding (Single (100) plane displayed and Nd not shown for clarity). Blue curves indicate hopping of mobile electrons between neighboring sites, which is suppressed between Al and O$_b$ due to the high energy of the unoccupied states on Al.

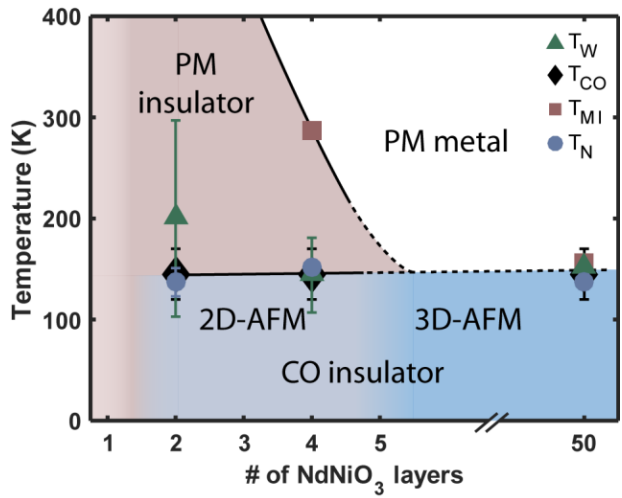

Fig. 9. Phase diagram of NdNiO$_3$ superlattices as a function of the number of confined NdNiO$_3$ layers, $m$, along with data from a 50 uc thick NdNiO$_3$ film (PM = paramagnetic, AFM = antiferromagnetic, CO = charge ordered). $T_{MI}$ (red squares) and $T_W$ (green triangles) are determined from transport, $T_{CO}$ is determined from XAS, and $T_N$ (blue circles) is determined from RSXS. Error bars denote uncertainty in fits to experimental data. Solid and dashed lines are guides to the eye.

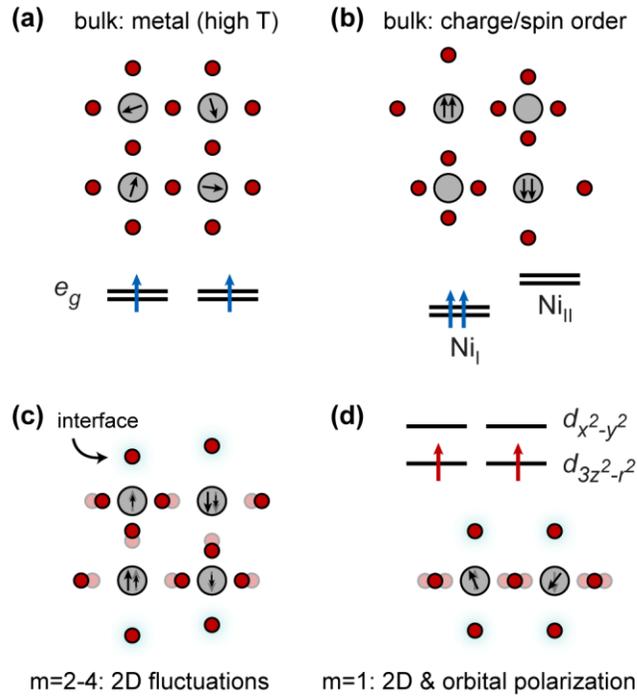

Fig. 10. Schematics of the atomic and electronic structure for (a) the high temperature, bulk metallic phase; (b) the low temperature bulk charge and spin ordered phase; (c) the m=2 superlattice; and (d) the m=1 superlattice. Ni and O are shown as grey and red circles, respectively. Orbital configurations ignore O $p$ states for clarity.

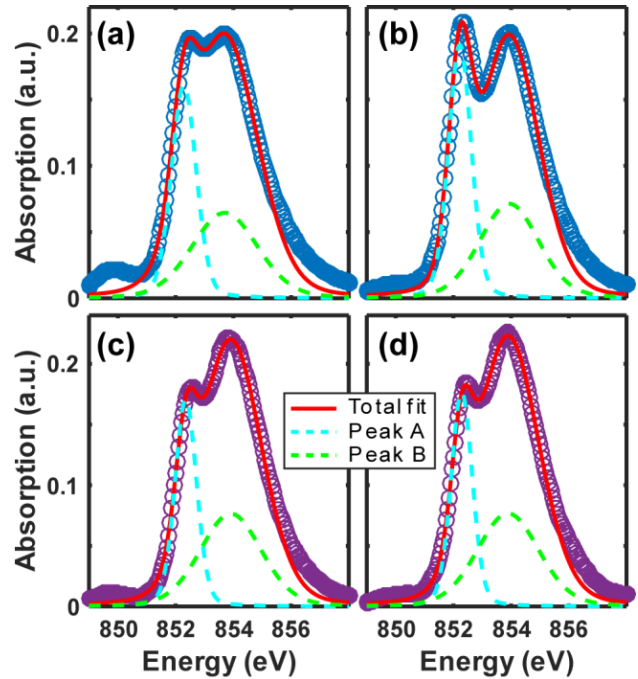

Fig. 11. Examples of measured Ni $L_3$ edge XAS spectra (circles), along with results of two-peak fitting (solid and dashed lines) using Eq. 9. Spectra shown are from the (a) 50 uc $NdNiO_3$ film at $T$ = 300 K and (b) 77 K, and from the (c) $m$=1 $NdNiO_3$/$NdAlO_3$ superlattice at $T$ = 300 K and (d) 77 K.

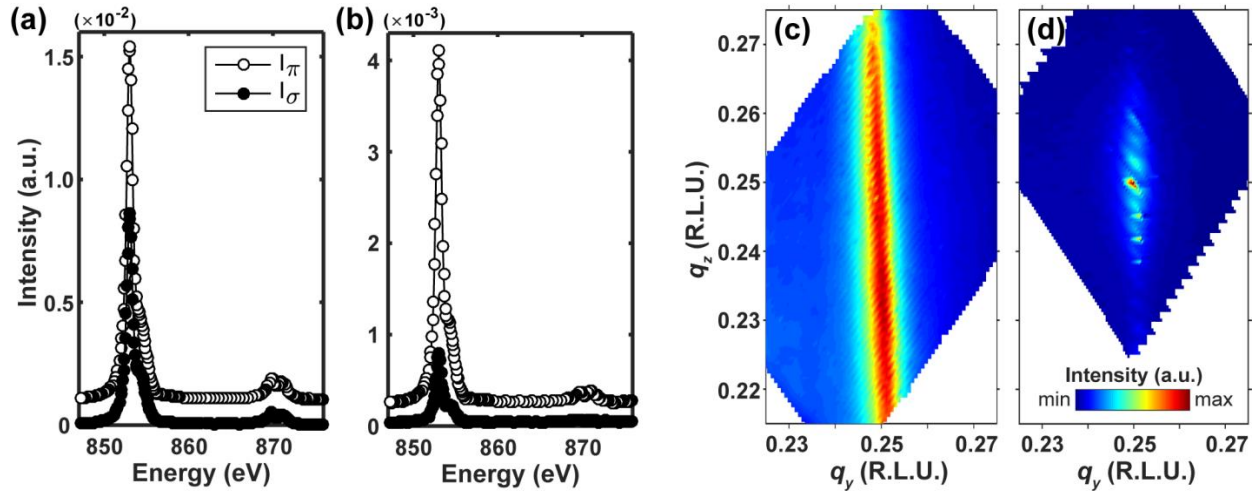

Fig. 12. (a,b) Energy dependence of $q$ = (¼,¼,¼) scattering intensity through the Ni $L$ edge for the (a) $m$=4 and (b) $m$=2 superlattice taken at $T$ = 30 K and $T$ = 15 K, respectively. (c,d) Reciprocal space map of magnetic scattering around $q$ = (¼ ¼ ¼) for the (c) $m$=4 superlattice and (d) 50 uc $NdNiO_3$ film taken at the Ni $L_3$ peak energy ($E$ = 853 eV) and $T$ = 30 K. Inset of (d) shows linear intensity scale for both maps is shown (max/min are different for each map).

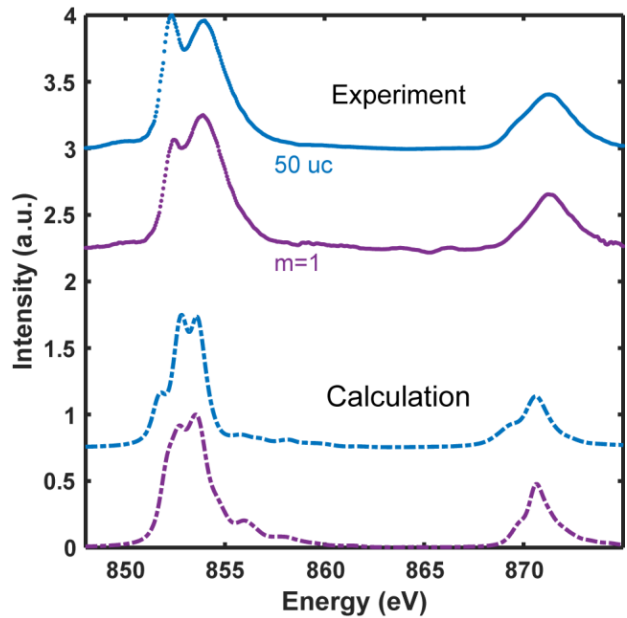

Fig. 13. XAS data measured at $T$=77K for $NdNiO_3$ and $m$=1 superlattice, along with calculated spectra (see text for details of calculations).

Table I. Electronic and structural parameters from DFT for bulk NdNiO$_3$ and NdNiO$_3$/NdAlO$_3$ superlattices. $\Delta_{e_g} = E(d_{3z^2-r^2}) - E(d_{x^2-y^2})$ is the Ni $e_g$ orbital energy splitting. The electron occupancies $n(O_a)$ and $n(O_b)$ are those of the in-plane and apical O $2p_\sigma$ orbitals that point along Ni-O bonds, and $n(O_c)$ is the occupancy of an O $2p_\sigma$ orbital in the AlO$_2$ plane of the NdAlO$_3$ layer adjacent to the interface (see Fig. 8). $d_{ap}/d_{inp}$ is the apical to in-plane Ni-O bond length ratio. Note that $n(O_b)$ and $d_{ap}/d_{inp}$ are averaged over the two apical O sites – closer and farther from Al – for the $m=2$ superlattice.

|  | $\Delta_{e_g}$ (eV) | $n(O_a)$ | $n(O_b)$ | $n(O_c)$ | $d_{ap}/d_{inp}$ |
|---|---|---|---|---|---|
| Bulk NdNiO$_3$ | -0.02 | 1.68 | 1.68 | --- | 1.01 |
| $m=2$ | -0.10 | 1.68 | 1.73 | 1.92 | 1.01 |
| $m=1$ | -0.26 | 1.68 | 1.78 | 1.92 | 1.01 |